\documentclass[fleqn,usenatbib]{mnras}

\usepackage{newtxtext,newtxmath}

\usepackage[T1]{fontenc}
\usepackage{ae,aecompl}
\usepackage{subfigure}

\usepackage{graphicx}	
\usepackage{amsmath}	
\usepackage{soul}
\usepackage{ulem}

\newcommand{\is}[1]{{#1}} 
\newcommand{\rb}[1]{{#1}} 

\newcommand{\rev}[1]{{#1}} 

\title[Compression of cosmic rays in SNRs]{Limits on compression of cosmic rays in supernova remnants}

\author[Sushch \& Brose]{
Iurii Sushch$^{1,2}$\thanks{E-mail: iurii.sushch@nwu.ac.za} and
Robert Brose$^{3}$
\\
$^{1}$Centre for Space Research, North-West University, 2520 Potchefstroom, South Africa\\
$^{2}$Astronomical Observatory of Ivan Franko National University of Lviv, Kyryla i Methodia 8, 79005 Lviv, Ukraine\\
$^{3}$Dublin Institute for Advanced Studies, Astronomy \& Astrophysics Section, 31 Fitzwilliam Place, D02 XF86 Dublin 2, Ireland
}
\date{Accepted 2023 February 19. Received 2023 January 23; in original form 2022 October 12}

\pubyear{2023}

\begin{document}
\label{firstpage}
\pagerange{\pageref{firstpage}--\pageref{lastpage}}
\maketitle

\begin{abstract}
The spectral shape of the gamma-ray emission observed for dynamically old supernova remnants that interact with molecular clouds triggered an exciting scenario of adiabatic compression and farther re-acceleration of Galactic cosmic rays (GCRs) in radiative shells of the remnants, which was extensively discussed and applied to various sources over recent years. Indeed, the observed gamma-ray spectrum from a number of remnants strongly resembles the expected spectrum of the gamma-ray emission from the compressed population of Galactic cosmic rays. In the following we discuss the feasibility of this scenario and show that it is very unlikely that compressed GCRs could produce sufficient amount of gamma-rays \rb{and that the observed spectral shape is putting strong limits on the allowed compression factors}. \rb{Further, absence of curvature in featureless power-law spectra of evolved supernova remnants at radio wavelengths is strongly disfavoring the compression scenario for electrons and hence for hadrons. Our calculations} show that the contribution of compressed electrons to the observed radio-flux could reach at most $\sim10\,$\%.
\end{abstract}




\section{Introduction}
\label{sec:intro}
Two decades of observations with the $Fermi$ LAT telescope has revealed a large population of gamma-bright supernova remnants \citep[SNRs;][]{2016ApJS..224....8A}. A significant fraction of this population constitute dynamically old SNRs, interacting with dense molecular clouds, whose gamma-ray emission can be confidently attributed to hadronic processes \citep{2013Sci...339..807A,2011MmSAI..82..747G,2016ApJ...816..100J,2019A&A...623A..86A, 2020MNRAS.497.3581D}. Another \rb{feature} that these remnants have in common is the shape of their gamma-ray spectrum which appears to be soft indicating a soft spectrum of the underlying proton poplulation with spectral indices of $\sim2.4-2.8$. Such a soft spectrum is expected for dynamically old SNRs, e.g. due to the combined effect of the decrease of the maximum energy and particle escape \citep{2019MNRAS.487.3199C, 2020A&A...634A..59B} and/or the decrease of the shock compression due to the propagation through the hot shocked wind of the progenitor star \citep{2022A&A...661A.128D}. The resemblance of the underlying proton spectrum with the spectrum of Galactic cosmic rays (GCRs) triggered also another exciting scenario in which pre-existing GCRs can be compressed and re-accelerated subsequently emitting gamma-ray radiation. \is{Such a mechanism is possible when the SNR shock is interacting with a dense material either in form of a large cloud or small very dense clumps that leads to formation of the radiative shell behind the shock front.} This \is{scenario} was proposed for a number of SNRs including the most prominent hadronic emitters W44 and IC~443 \citep{2010ApJ...723L.122U, 2015ApJ...806...71L, 2016A&A...595A..58C,2014ApJ...784L..35T, 2015ApJ...800..103T, 2019MNRAS.482.3843T} \is{closely following the idea proposed by \citet{1982ApJ...260..625B}}. However, it was shown recently that some of models suffer from strong inconsistency connected to the unrealistically large amount of gas required in the shell of the remnant \citep{2020MNRAS.497.3581D}. Indeed, to match the observed flux, it is required that the strongly compressed shell which consists of the crushed cloud material covers a large fraction of the SNR volume, which results in more material in the shell than the SNR can possibly acquire from the ambient medium during its evolution.  On the other hand, it should also be stressed, that under the assumption that SNRs are sources of GCRs, it is expected that the spectrum of particles released by SNRs into the medium follows a power law with the spectral index of $\sim2.4$ and hence, the downstream spectrum of \is{particles accelerated at shocks of dynamically old SNRs must be} softer than that.

Because compression/re-acceleration models are still widely discussed and promoted within the community we would like to follow-up and explore further their feasibility. In the following we will consider only compression of GCRs as the re-acceleration process does not strongly impact the normalization of the resulting gamma-ray spectrum but rather shifts it in energy.

\newpage

\section{Adiabatic compression of GCR protons and electrons}

 The basic idea of the compression process is that the interaction of the SNR shock with the dense \is{environment} results in the formation of the radiative shell behind the shock front. The material behind the shock is adiabatically compressed to very high densities potentially boosting any pion-decay generated gamma-ray emission and also radio emission by enhancing the magnetic-field strength. The adiabatic compression of the pre-existing ambient GCRs in the radiative shell enhances the CR spectrum both energizing particles and increasing the normalization of the spectrum. The compressed CR spectrum can then be expressed as \citep{2010ApJ...723L.122U, 1982ApJ...260..625B}
\begin{equation}
    \label{eq:speccompression}
    n_\mathrm{comp}(p) = \xi^{2/3}n_\mathrm{GCR}(\xi^{-1/3}p)\text{ , }
\end{equation}
where $n_\mathrm{GCR}(p)$ is the density of GCRs as a function of momentum and $\xi \equiv n_\mathrm{shell}/(rn_0)$ is the adiabatic compression ratio, with $n_\mathrm{shell}$ the density of the cooled gas in the shell, $n_0$ the density of the ambient medium (cloud), and $r$ the shock compression ratio. The interaction of compressed GCRs with the compressed cloud material of high density and high magnetic field in the shell can potentially result in substantial gamma-ray and radio emission. 

\subsection{Galactic protons}
For the proton CR spectrum we adopt the approximation of the observed proton flux proposed by \citet{2016Ap&SS.361...48B} imposed with a spectral hardening at higher energies \citep{2011Sci...332...69A, 2015PhRvL.114q1103A}: 
\begin{align}
J_\mathrm{GCR}(E) = &0.3719 \frac{E^{1.03}}{\beta^2} \left(\frac{E^{1.21} + 0.77^{1.21}}{1+0.77^{1.21}}\right)^{-3.18} \\ \nonumber &\times\left[1+\left(\frac{E}{335}\right)^{\frac{0.119}{0.024}}\right]^{0.024}  \left[\mathrm{GeV}^{-1} \mathrm{cm}^{-2} \mathrm{sr}^{-1} \mathrm{s}^{-1}\right]      
\end{align}
where $E$ is the kinetic energy of proton in GeV and $\beta$ is the proton velocity in $c$. The number density of CRs as a function of momentum is given then by
\begin{equation}
    n_\mathrm{GCR}(p) = \beta c \, n_\mathrm{GCR}(E) = 4\pi J_\mathrm{GCR}(E) 
\end{equation}
and the compressed spectrum can be found using Eq.~\ref{eq:speccompression}.

The level of the resulting gamma-ray emission is completely determined by the compression ratio and the amount of the target material in the shell. For a specific SNR with known size and density of the ambient medium (cloud) this simplifies to just two parameters: the total compression ratio $\chi \equiv n_\mathrm{shell}/n_0 = \xi r$ and volume filling factor $f = V_\mathrm{shell}/V_\mathrm{SNR}$, the ratio of the volume of the shell to the volume of the remnant. It is however often ignored that these two parameters are interdependent and their parameter space is strongly constrained.

\subsection{Galactic electrons}

The Galactic electron spectrum is well described by a power-law with an power-law index of $s_1=3.04$ at higher energies. However, the spectrum shows a smooth transition to a spectrum with $s\approx1$ at lower energies \citep{2011MNRAS.416.1152J}. This spectral shape can be described by a log-parabola at low energies that transitions to a power-law at higher energies,
\begin{align}
N_\mathrm{e}(E) = 
        \begin{cases}
            \frac{N_1}{E}\exp\left(-\frac{\log^2({E}/{E_\mathrm{B}})}{\sigma}\right)&\mathrm{ for }E\leq E_\mathrm{B}\\
            N_2E^{-s} &\mathrm{ for }E>E_\mathrm{B}
        \end{cases}\label{Eq:BackgroundELs}\mathrm{.}
\end{align}
A fit to the electron spectra given in \cite{2011MNRAS.416.1152J} for a galactocentric radius of $6.5\,$kpc with expression (\ref{Eq:BackgroundELs}) yields $s = 3.04$ and $E_\mathrm{B}=$ $5\,$GeV. 
These spectra are compatible with direct observations of the electron spectra in the local ISM by Voyager 1, which also show spectra harder than $s=2.0$ at low energies \citep{2016ApJ...831...18C}.

\subsection{Constraint imposed by the available cloud material}

The radiative shell behind the shock front where adiabatic compression takes place consists of the cloud material overran by the shock. This imposes a hard upper limit on the total number of particles in the shell for a current given volume of the SNR
\begin{equation}
    N_\mathrm{max} = V_\mathrm{SNR} n_0,
\end{equation}
where $V_\mathrm{SNR} = 4/3 \pi R_\mathrm{sh}^3$, $R_\mathrm{sh}$ is the shock radius, and $n_0$ is the particle density of the cloud. This upper limit follows from a simple consideration that if the SNR has been always expanding into the cloud throughout its evolution that is the maximum possible number of particles that can be located behind the shock.

Now, the total number of particles in the shell can be expressed as
\begin{equation}
    \label{eq:Nshell}
    N_\mathrm{shell} = f V_\mathrm{SNR} n_\mathrm{shell} = f V_\mathrm{SNR} \chi n_0
\end{equation}

From this immediately follows a condition on the product of the volume factor and the total compression ratio:
\begin{equation}
    f\chi \leq 1 \label{eq:chi_basic}
\end{equation}
\is{It should be noted that the above constraint is independent of the structure of the medium, size of clouds, their density and the amount of volume that they occupy. It follows solemnly from a basic condition that the crushed shell cannot possibly contain more particles than available in the ambient medium confined by the current size of the SNR. This condition is not fulfilled e.g. in \citet{2016A&A...595A..58C} for W44 \rev{where the model requires the filling factor of $f=0.14$ and the total compression ratio of $\chi = 50$\footnote{\rev{Note, \citet{2016A&A...595A..58C} uses the surface filling factor in their calculations and the relation of filling factor $f$ that is defined the same way as in this work and surface filling factor $\xi$ is shown on page 5 of \citet{2016A&A...595A..58C}. $f=0.14$ corresponds to $\xi = 0.55$ required by their model. $\chi=50$ follows from the required density of the crushed shell of 10,000 cm$^{-3}$ with the adopted cloud density of 200 cm$^{-3}$}}},  while in \citet{2010ApJ...723L.122U} the model operates at $10-20$\,\% level of this hard upper limit.}

In reality, the product $f\chi$ should be much lower than unity as considered SNRs interact with clouds only during a certain fraction of their evolution and the interaction does not cover the whole shock surface. Moreover, the radius of the SNR expanding into the dense cloud medium throughout its evolution is expected to be much smaller than observed for specific SNRs. In fact, the radius of the SNR at the beginning of the radiative or pressure-driven-snowplough (PDS) stage can be expressed as \citep{1988ApJ...334..252C}
\begin{equation}
    R_\mathrm{PDS} = 1.2 \left[\frac{n_0}{300\,\mathrm{cm}^{-3}}\right]^{-3/7}\left[\frac{E_\mathrm{SN}}{10^{51}\,erg}\right]^{2/7}\zeta_\mathrm{m}^{-1/7}\,\mathrm{pc},
\end{equation}
where $E_\mathrm{SN}$ is the supernova explosion energy and $\zeta_\mathrm{m}$ is the metallicity factor, $\zeta_\mathrm{m}=1$ for solar abundances. This radius can be considered as an upper limit on the distance that the SNR would expand into the cloud at any stage of its evolution, which would provide us with a more constraining limit on the $f\chi$ product. Assuming that the whole surface of the SNR shock expanded into the cloud for the distance of $\Delta R$, the total volume of the cloud that shock interacted with can be written as
\begin{equation}
    V_\mathrm{cloud} = \frac{4}{3}\pi(R_\mathrm{sh}^{3} - (R_\mathrm{sh}-\Delta R)^3) \approx 4\pi R_\mathrm{sh}^{2} \Delta R\left(1 - \frac{\Delta R}{R_\mathrm{sh}}\right)
\end{equation}
Hence, the total number of particles that can be accumulated in the shell is 
\begin{equation}
    N = V_\mathrm{cloud}n_0 = 4\pi R_\mathrm{sh}^{2} \Delta R\left(1 - \frac{\Delta R}{R_\mathrm{sh}}\right) n_0
\end{equation}
Combining this with Eq.~\ref{eq:Nshell} one gets 
\begin{equation}
    f\chi = 3 \frac{\Delta R}{R_\mathrm{sh}}\left(1 - \frac{\Delta R}{R_\mathrm{sh}}\right)
\end{equation}
or 
\begin{equation}
    \label{eq:upperlimit}
    f\chi \leq 3 \frac{R_\mathrm{PDS}}{R_\mathrm{sh}}\left(1 - \frac{R_\mathrm{PDS}}{R_\mathrm{sh}}\right)\text{ . } 
\end{equation}

\is{The large radii of observed SNRs are difficult to accommodate within the assumption of an expansion into a uniform medium with a high density. This problem can be overcome using the assumption of a clumpy medium, where dense but small clumps are responsible for the creation of the crushed shell but are dynamically unimportant. This assumption would also allow for the high shock velocities during interaction with dense clumps even at late stages of evolution. The important parameter that characterizes such a medium is the filling factor $\phi$ which expresses the fraction of the volume occupied by the clumps. It is clear that the filling factor cannot be too large to justify the assumption that the clumps are dynamically unimportant. In the following, we consider an upper limit on the filling factor of $\phi\leq0.1$ which follows from hydrodynamic simulations \citep[see e.g.][]{2017ApJ...846...77S} and still reflects a quite optimistic scenario. It should be noted, however, that even for low filling factors the SNR shock will expand slower than in the uniform intercloud medium due to conductive evaporation of clumps embedded in the hot gas behind the shock and such evolution would be accompanied by significant thermal X-ray emission \citep{1991ApJ...373..543W,2017ApJ...846...77S}.

For the most optimistic scenario, where the SNR expands into the clumpy medium throughout its \rev{whole} evolution the filling factor of clumps $\phi$ can be expressed by the means of the defined above filling factor of the crushed shell $f$ and total compression ratio $\chi$ as
\rev{
\begin{equation}
    \phi = \frac{V_{\mathrm{clumps}}}{V_\mathrm{SNR}} =\frac{V_{\mathrm{shell}}}{V_\mathrm{SNR}} \frac{V_{\mathrm{clumps}}}{V_\mathrm{shell}} =  f\chi
\end{equation}
}
and therefore implies a condition
\begin{equation}
    \label{eq:upperlimit_clumps}
    f\chi\leq0.1
\end{equation}
to justify the neglect of clumps in the shock dynamics. Note, that this condition is not satisfied in \citet{2010ApJ...723L.122U} where best-fit models require the clumps filling factor of $\phi \sim 0.2-0.4$\footnote{Note, that $\phi$ is not the same as the filling factor $f$ in \citet{2010ApJ...723L.122U} which is also defined as the ratio of the volume occupied by clumps to the volume of the SNR, but it is assumed that the SNR interacts with the clumpy medium only during a half of its age.} and hence clumps cannot be considered as dynamically unimportant.

In the following, we probe the non-thermal emission from compressed CRs in both scenarios, i.e. the expansion of the SNR into one large cloud (condition~\ref{eq:upperlimit}) and the expansion into a clumpy medium (condition~\ref{eq:upperlimit_clumps}). In both cases we use the same radius of the SNR ignoring differences in the SNR evolution and focusing on testing the requirement of an sufficient amount of cloud material. This would naturally result in a larger age and a lower shock velocity in the one-large-cloud scenario.
}

\subsection{Constraints on the total compression ratio}

Assuming the magnetic field in the cloud, $B_0$, is turbulent, the magnetic field in the compressed shell can be expressed by
\begin{equation}
    B_\mathrm{shell} = \sqrt{\frac{2\chi^2+1}{3}} B_0
\end{equation}
or 
\begin{equation}
    B_\mathrm{shell} \approx \sqrt{2/3}\chi B_0    
\end{equation}
for large $\chi$. Now, following the assumption that compression due to radiative cooling downstream is limited by the magnetic pressure one can estimate the total compression ratio by equating the magnetic pressure with the shock ram pressure \citep{1982ApJ...260..625B, 2010ApJ...723L.122U}:
\begin{equation}
    \chi \simeq 9.4 \left[\frac{n_0}{1\,\mathrm{cm}^{-3}}\right]^{1/2}\left[\frac{B_0}{1\,\mu\mathrm{G}}\right]^{-1} \left[\frac{v_\mathrm{sh}}{10^6\,\mathrm{cm/s}}\right]
    \label{eq:compression}
\end{equation}

The study on deducing magnetic field strengths in molecular clouds from Zeeman observations by \cite{2010ApJ...725..466C} indicates roughly constant magnetic fields of $B_0=10\,\mu$G in clouds with densities of $n_0\lesssim 300\,\text{cm}^{-3}$ with a power-law increase for denser clouds. The generalized empirical model for the maximum magnetic field strength in the cloud can be expressed as \citep{2010ApJ...725..466C}:
\begin{equation}
    B_0 = \begin{cases}
        10\,\mu\mathrm{G}, & n_0\leq300\,\text{cm}^{-3}\\
        10\,\mu\mathrm{G}\left(\frac{n_0}{300\,\text{cm}^{-3}}\right)^{0.65} & n_0>300\,\text{cm}^{-3}
    \end{cases}\label{eq:B_Crutcher}
\end{equation}

Adopting this result into Eq.~\ref{eq:compression} implies that the compression ratio could be as small as
\begin{equation}
    \chi \gtrsim \begin{cases}
        16.3 \left[\frac{n_0}{300\,\mathrm{cm}^{-3}}\right]^{1/2} \left[\frac{v_\mathrm{sh}}{10^6\,\mathrm{cm/s}}\right] & n_0\leq300\,\text{cm}^{-3}\\
        16.3 \left[\frac{n_0}{300\,\mathrm{cm}^{-3}}\right]^{-0.15} \left[\frac{v_\mathrm{sh}}{10^6\,\mathrm{cm/s}}\right] & n_0>300\,\text{cm}^{-3}
    \end{cases}
\end{equation}
While for high-density clouds this should be considered as a lower limit, for low-density clouds it basically reflects the estimate of the compression ratio as for many clouds the magnetic field is found to be at the level of $\sim10\,\mu$G \citep{2010ApJ...725..466C}.

The shock velocity is strongly dependent on the density of the cloud. For the constant density of the ambient medium of $1$~cm$^{-3}$ an SNR reaches the radiative stage of evolution when its shock velocity is a few hundred km/s \citep[see e.g.][]{1988ApJ...334..252C}. The shock velocity in the dense cloud could be considerably smaller than that. Generally, the shock velocity in the cloud can be approximated by \citep{1975ApJ...195..715M}
\begin{equation}
    v_\mathrm{sh} \sim v_\mathrm{sh}^\prime \sqrt{\frac{n_0}{n_0^\prime}},
\end{equation}
where $v_\mathrm{sh}^\prime$ and $n_0^\prime$ are the shock velocity and the density of the medium before the interaction. Combining this with Eq.~\ref{eq:compression} one can conclude that the total compression ratio $\chi$ does not strongly depend on the density of the cloud. Moreover, it should decrease for densities larger than $\sim300$~cm$^{-3}$ due to the increase of the magnetic field. The shock velocity in the cloud for typical values should be around $\sim10-100$ km/s resulting in compression ratio of $\chi\sim10-100$. On the other hand, the shock velocity has to be larger than the speed of sound in the cloud (typically about 10~km/s) at least by a factor of two to ensure that the shell is not disrupted and merged with the medium. This means that the compression ratio should also be at least a few tens.

\begin{figure}
    \centering
    \includegraphics[width=\linewidth]{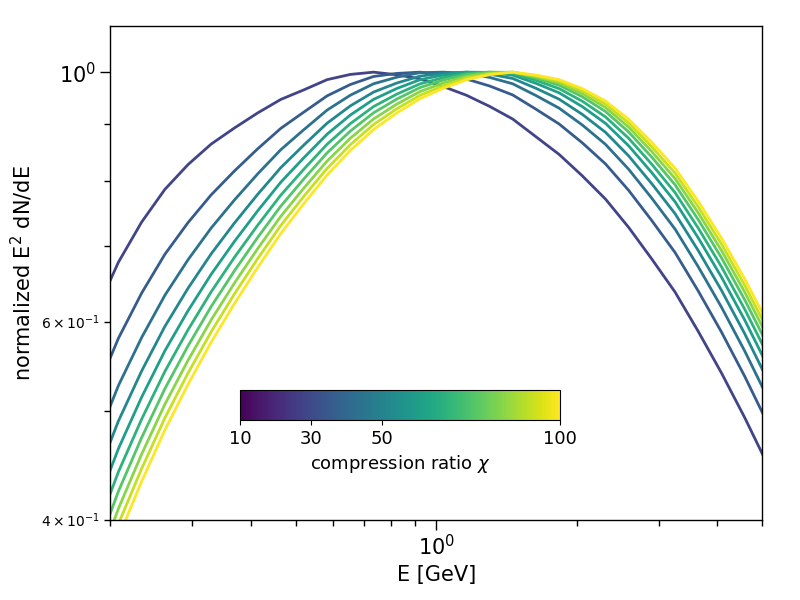}
    \caption{Normalized spectral energy distribution of the gamma-ray emission produced by the compressed GCR protons for different values of the compression ratio $\chi$. Curves are plotted for $\chi$ in the range from 10 to 100 with the step of 10. Gamma-ray emission is calculated as described in Section~\ref{sec:gammaray}}
    \label{fig:energy_constraint}
\end{figure}

Another constraint on the compression ratio follows from the observed spectral shape of the gamma-ray emission for these dynamically old interacting SNRs, for which the compression scenario could be applicable. Most of them exhibit a peak in the energy flux at $\lesssim1$~GeV with maybe one outlier, G$349.7+0.2$, which seems to have a peak at around 2~GeV \citep[see figure 1 in][]{2019MNRAS.482.3843T}. Such a low peak can be reproduced only for moderate values of the compression ratio as the adiabatic compression also energizes particles, and for too high values of $\chi$ the peak in the gamma-ray spectrum would move to higher energies. In Figure~\ref{fig:energy_constraint} we show the normalized spectral energy distribution of the resulting gamma-ray radiation for different values of $\chi$. The energy of the peak is 0.7~GeV for $\chi = 10$, and it shifts to 1 GeV already for $\chi = 30$. For $\chi = 100$ the peak is located at 1.5~GeV with a quite steep decrease below that energy. This suggests that from the observational point of view the compression ratio cannot be larger than a few tens. Note, that this simple analysis does not take into account additional reacceleration of GCRs, which would bring the constraint on $\chi$ to even lower values.

As one can see the value of the compression ratio is constrained on both sides through different consideration to a few tens, which strongly limits the parameter space and leaves little room to maneuver.

\section{Gamma-ray emission}
\label{sec:gammaray}

To calculate the gamma-ray emission we use the post-processing radiation routine of the \textsc{RATPaC} code \citep{2012APh....35..300T, 2013A&A...552A.102T, 2016A&A...593A..20B, 2018A&A...618A.155S} designed for numerical simulations of particle acceleration in SNRs. The module to calculate gamma-ray radiation from pion decays relies on  Monte-Carlo event generators, namely {\sc DPMJET}-III \citep{roesler2001monte} and UrQMD \citep{1998PrPNP..41..255B, bleicher1999relativistic}, for the calculation of inelastic cross sections and differential production rates of secondary particles produced in nuclei collisions \citep{2020APh...12302490B}. As target particles for hadronic interactions we take into account both H and He with a typical composition H$:$He$=10:1$.  

\begin{figure}
    \centering
    \begin{subfigure}
        \centering
        \includegraphics[width=\linewidth]{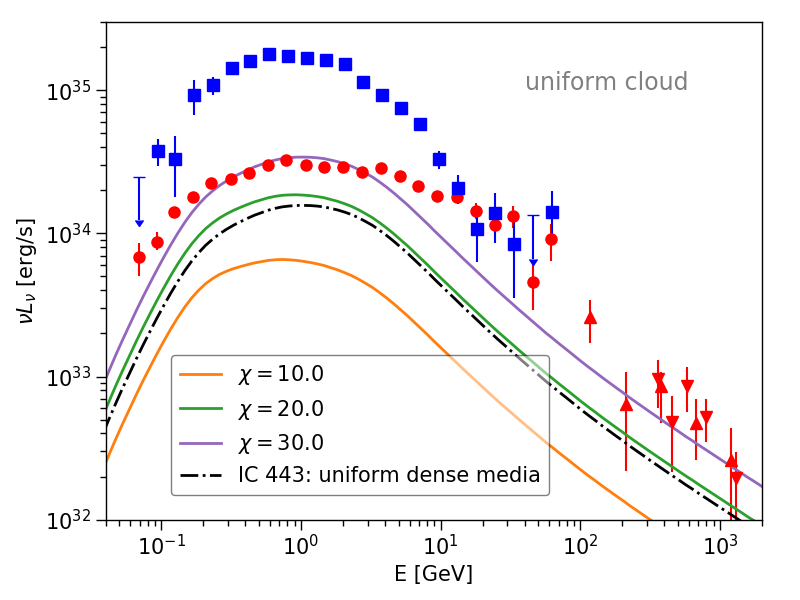}
    \end{subfigure}
    \begin{subfigure}
        \centering
        \includegraphics[width=\linewidth]{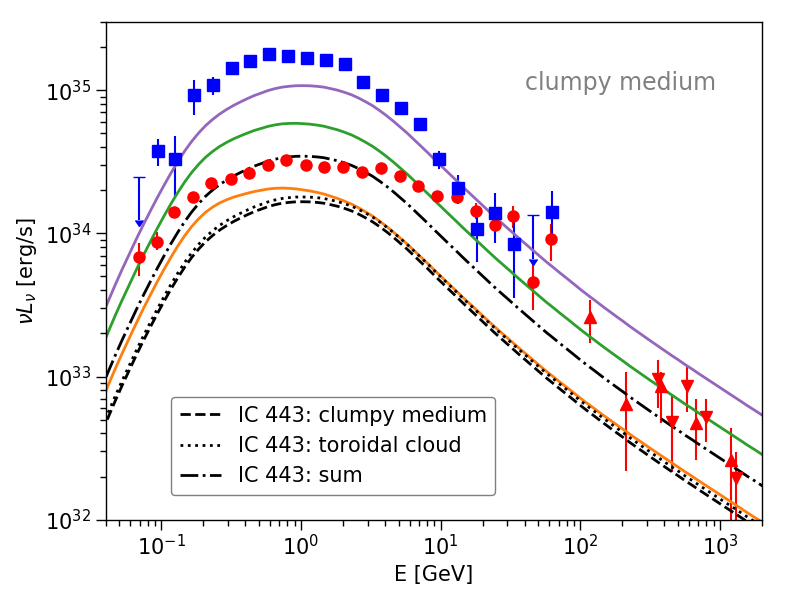}
    \end{subfigure}
    \caption{Solid curves show upper limits on the spectral gamma-ray luminosity resulted from the compression of GCR protons in dynamically old SNRs interacting with \is{a uniform cloud (top panel; satisfies the condition in Eq.~\ref{eq:upperlimit}) and clumpy medium (bottom panel; satisfies the condition in Eq.~\ref{eq:upperlimit_clumps})} for different values of the compression ratio. The shock radius is set to 10~pc \is{for both scenarios.} The density of the \is{uniform} cloud is set to $300\,\mathrm{cm}^{-3}$ \is{while the density of clumps is set to $1000\,\mathrm{cm}^{-3}$. In both scenarios the density of the intercloud medium is set to $1\,\mathrm{cm}^{-3}$ which roughly results in the average density of $\sim100\,\mathrm{cm}^{-3}$ for both cases.} Dash-dotted lines illustrate models constructed specifically for the IC~443 SNR  \is{under assumption of the uniform dense medium (top panel) and complex clumpy medium (bottom panel; see the Appendix for detailed description). Dashed and dotted lines in the bottom panel show individual components of the second model. } \is{Red markers show} the observed spectrum of the IC~443 SNR with Fermi LAT \citep[circles;][]{2013Sci...339..807A}, MAGIC \citep[upward triangles][]{2007ApJ...664L..87A}, and VERITAS \citep[downward triangles;][]{2009ApJ...698L.133A}. The Fermi LAT data for the W44 SNR \citep{2013Sci...339..807A} is shown with blue squares.}
    \label{fig:gammaray_lum}
\end{figure}

In Figure~\ref{fig:gammaray_lum} we show upper limit curves \is{with solid lines} for the gamma-ray spectral luminosity \is{in the uniform cloud scenario (top panel) and clumpy medium scenario (bottom panel)} for three values of the compression ratio \is{indicated with different colors}. We set the radius of the shock to 10~pc. The density of the cloud \is{in the uniform cloud scenario is set} to $300\,\mathrm{cm}^{-3}$ which reflects the turning point above which observationally magnetic field in clouds starts to increase with the density. \is{In the clumpy medium scenario the density of clumps is set to $1000\,\mathrm{cm}^{-3}$. In both scenarios the density of the intercloud medium is fixed at $1\,\mathrm{cm}^{-3}$, i.e. the average density is about $100\,\mathrm{cm}^{-3}$ for both cases.} The volume filling factor \is{of the crushed shell $f$} is set to the upper limit value as follows from Eq.~\ref{eq:upperlimit} \is{for the uniform cloud scenario and Eq.~\ref{eq:upperlimit_clumps} for the clumpy medium. The volume filling factor of clumps in the medium is set to $\phi = 0.1$.} The data points show the \is{spectra of the two most famous and studied hadronic SNRs, IC~443 (Fermi LAT \citep[red circles;][]{2013Sci...339..807A}, MAGIC \citep[red upward triangles][]{2007ApJ...664L..87A}, and VERITAS \citep[red downward triangles;][]{2009ApJ...698L.133A}}) and W44 (Fermi LAT \citep[blue squares;][]{2013Sci...339..807A}). 
\is{IC~443 is the least luminous in gamma-rays} among other aged SNRs \is{while W44 falls roughly in the middle of the distribution based on the gamm-ray luminosity} \citep{2019MNRAS.482.3843T}. IC~443 also exhibits a slightly higher peak energy, close to 1~GeV, that allows to vary $\chi$ up to 30. For other SNRs except G$349.7+0.2$ the peak energy is lower than that, constraining the compression ratio even more. G$349.7+0.2$, on the other hand, exhibits a very high luminosity which would be hard to reach even for higher values of $\chi$. It can be seen \is{in the top panel of Fig.} \ref{fig:gammaray_lum} that for \is{the scenario of the large uniform cloud covering the whole surface of the forward shock the derived upper limit on gamma-ray emission from the compression of GCR protons can barely reach the gamma-ray luminosity of IC~443 and falls almost an order of magnitude below the the gamma-ray luminosity of W44 for the most optimistic case. The situation is better for the clumpy medium scenario (bottom panel of Fig.~\ref{fig:gammaray_lum}) where derived upper limits do not exclude the GCRs model for IC 443 but still fall below the observed gamma-ray luminosity of W44.
For SNRs more luminous than W44 the contribution of compressed GCR should be negligible.} Moreover, any potential additional reacceleration of CRs would shift the peak towards higher energies even more limiting the range of allowed values of $\chi$. \is{Therefore,} the failure of the adiabatic compression in providing enough gamma-ray flux around the peak-energy, means that reacceleration models based on the compressed CR spectrum will fail too.

For comparison we also show two models constructed specifically for the IC~443 SNR which follow observational properties of the remnant. \is{The first model assumes uniform dense media with different densities on two sides of the remnant (Fig.~\ref{fig:gammaray_lum} top panel) and second model considers a clumpy medium in the northern hemisphere of the remnant and additional interaction with a very dense toroidal molecular cloud}. The models are \is{further} described in the Appendix. Both models \is{optimistically} assume the compression ratio of $\chi=30$ and fall below \is{respective upper limits. The 'uniform dense media' model fails to explain the observed IC 443 flux while the 'clumpy medium' model can roughly fit the data indicating that indeed there might be some contribution from compressed GCRs in the case of IC 443. } 
\is{However, it should be stressed }
here again that fluxes \is{in both models} are derived under optimistic conditions and can be considered as upper limits of the gamma-ray emission from compressed GCRs in IC~443\is{, which is probably the best candidate for such a model given its low gamma-ray luminosity and a very dense and complex medium.}

\section{Radio emission}

The compressed Galactic electrons together with the compressed magnetic fields in the SNR-shell will boost the radio-synchrotron emission from these objects. In fact, it is hard to explain how CR protons shall not be accelerated by the SNR shock, so that compression is the main driver of the gamma ray emission, while electrons get freshly accelerated. This is especially constraining since radio-emitting electrons and protons responsible for the gamma-ray emission have approximately the same energy. 

We use the Galactic electron spectrum as parameterized in Eq.~\ref{Eq:BackgroundELs} and compressed according to Eq.~\ref{eq:speccompression}. We then use the emission routines of \textsc{RATPaC} code \citep{2012APh....35..300T, 2013A&A...552A.102T, 2016A&A...593A..20B, 2018A&A...618A.155S} to calculate the synchrotron emission that can be expected from the compressed shell.  

\subsection{Spectral break}
The spectral turnover of the Galactic electron spectrum below $E_\text{B}=5\,$GeV leads to a spectral break in the synchrotron emission from the compressed electrons. The radio break-frequency $\nu_\text{B}$ can be calculated according to 
\begin{align}
    \nu_\text{B} &= 16\text{\,MHz}\left(\frac{\sqrt{2/3}\chi B_0}{\mu\text{G}}\right)\left(\frac{\xi^{1/3} E_\text{B}}{\text{GeV}}\right)\text{.}
\end{align}

As a result, the radio-spectra shows a turnover from a radio-spectral index of $\alpha=0$ to $\alpha=1$ below $\nu_\text{B}=40\,$GHz for $\chi=30$ and $B_0=10\,\mu$G. For a lower compression ratio of $\chi=10$, the transition happens below $\nu_\text{B}=9\,$GHz.

The fact that particularly evolved SNRs show remarkably featureless radio-spectra from low to high energies \citep[e.g.][]{2007A&A...471..537C} comprehends strong evidence against a compression-only origin of the Radio emission.

\subsection{Radio flux}
We calculated the radio flux from compressed electrons using the parameters from section \ref{sec:gammaray}. \rb{Here, we also consider two distinct scenarios, one assuming only a homogeneous medium and one assuming the presence of additional dense clumps. For the magnetic field we follow the upper-limits given by \cite{2010ApJ...725..466C} given in Eq.~\ref{eq:B_Crutcher}.} Note, that usage of Eq.~\ref{eq:upperlimit} for synchrotron radiation does not necessarily provide an upper limit on the emission as the emission does not depend on the number target particles in this case. 

\begin{figure}
    \centering
    \includegraphics[width=0.49\textwidth]{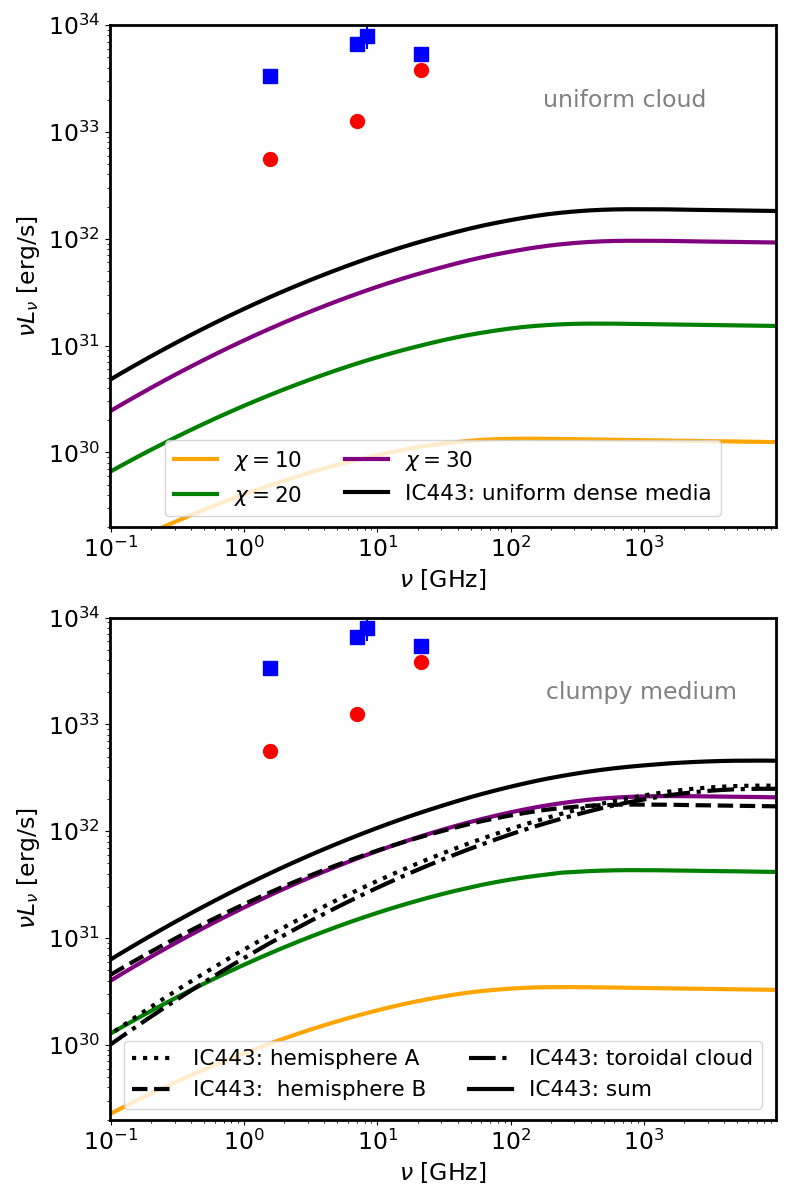}
    \caption{Solid \is{colored} curves show upper limits on the Radio-luminosity from a population of compressed electrons in an dynamically old and interacting SNR. \rb{The upper panel shows the emission for an ambient density of $300\,\text{cm}^{-3}$ and the lower panel for a $1\,\text{cm}^{-3}$ medium and $1000\,\text{cm}^{-3}$ clumps with a volume-filling factor of \is{$\phi=0.1$}. The shock-radius is set to 10pc in both cases. Red \is{circles} and \is{blue squares} indicate the observed \is{radio} luminosities for W44 and IC~443 respectively \citep{2017MNRAS.470.1329E, 2019MNRAS.482.3857L}. The black lines \is{illustrate} dedicated models for IC~443 \is{(see Appendix)} \is{including distinct components as described in the legend.}}} 
    \label{fig:W44_Radio}
\end{figure}

Figure \ref{fig:W44_Radio} shows a comparison of the expected radio-luminosity with the luminosities of W44 and IC~443. It is evident that the radio-luminosity from the compressed electrons is well below the observed radio-luminosity for both remnants \rb{and both considered scenarios}, contributing at most on a level of $\approx10\,$\%. The dedicated model\rb{s for IC~443 show slightly higher fluxes compared to the generic models on account of the compressed field in the clumps and the dense cloud} (see also Appendix \ref{app:models}). \rb{It should be noted here, that the magnetic-field scaling derived by \cite{2010ApJ...725..466C} constitutes an upper limit on the field in the cloud and clumps and so does consequently our predicted radio-emission.}
 
The underprediction of the radio-flux from compressed electrons strongly indicates that additional electrons need to be accelerated in the SNRs. These additional electrons can either be accelerated before the SNR-cloud interaction or continuously during the cloud-interaction. These freshly accelerated electrons would need to reach at least energies of a few GeV. In this case, however, also protons should be accelerated to at least those energies and would contribute to the gamma-ray emission.
 
Whereas the compression of Galactic electrons falls short in providing enough radio-luminosity in general, the situation might be different for other radio-sources. Recently, it was shown that non-thermal radio emission for the stellar bow-shock BD43 can be well explained by the compression of Galactic electrons in this system \citep{2022A&A...663A..80M}. However, in these systems, the shock-velocity is never exceeding a few tens of km/s, while SNRs experience considerably higher shock-velocities during the initial phases of their evolution. Here, as already noted in section \ref{sec:gammaray}, the fresh acceleration of electrons and protons in the initial phases of the remnants evolution can provide enough gamma-ray and radio flux, while the interaction with the dense cloud enhances particle escape and shapes the observed soft spectra \citep{2019MNRAS.487.3199C, 2020A&A...634A..59B, 2021A&A...654A.139B}.

\section{Particle confinement}
A consequence of condition in Eq.~\ref{eq:chi_basic} is, that the thickness of the compressed shell $D_\text{shell}$ can be written as
\begin{align}
    D_\text{shell} = R_\text{sh}\left( 1-\sqrt[3]{1-\frac{1}{\chi}}\right) \text{ . } \label{eq:dR}
\end{align}
This length-scale can be compared to the mean free path of particles in order to investigate if they can be confined by compression in the shell or if they escape the compression-process diffusely. The mean free path $L$ is roughly given by
\begin{align}
    L =& \frac{D(E)}{c} \approx A\cdot10^{19}\left( \frac{E}{\text{10 GeV}}\right)^{1/3}\left( \frac{\sqrt{2/3}\chi B_0}{3\,\mu\text{G}}\right)^{-1/3}\text{ cm, } \label{eq:MFP}
\end{align}
where $A$ is a numerical factor describing the suppression of the diffusion around the SNR. A value of $A=1$ implies the Galactic diffusion coefficient, while $A=10^{-2}$ is suppressed diffusion as observed around gamma-ray pulsars \citep{2017Sci...358..911A} or as derived for SNRs \citep{2010ApJ...712L.153F,2021A&A...654A.139B}. Equations \ref{eq:dR} and \ref{eq:MFP} can be combined to obtain the energy up to which particles can be confined,
\begin{align}
    E &= 1\left(\frac{R_\text{sh}}{A\cdot10^{19}\,\text{cm}}\right)^3\left(\frac{\sqrt{2/3}\chi B_0}{3\,\mu\text{G}}\right)\left( 1-\sqrt[3]{1-\frac{1}{\chi}}\right)^3\text{ TeV.}\label{eq:E_conf}
\end{align}
For a compression ratio of $\chi=30$ and $A=1$, Eq.~\ref{eq:E_conf} yields a energy of $3.4\,$GeV. If the diffusion coefficient is suppressed by a factor of 10, the energy-limit reaches $\approx3\,$TeV, enough to potentially compress CRs at the gamma-ray emitting energies. However, while the reduction of the diffusion coefficient is normally provided by escaping high-energy CRs, no such escaping particles are expected in the compression scenario.

Additionally, Eq.~\ref{eq:E_conf} indicates, that higher compression ratios yield a lower confinement energy as the decrease in the shell-thickness wins over the increase in the magnetic field. This, again, makes scenarios of very high compression factors unlikely.

\section{Conclusions}
\label{sec:conclusions}
 
The amount of cloud material that can be possibly swept up by the SNR shock as well as constraints on the compression ratio strongly limits the potential gamma-ray luminosity generated by compressed GCR protons. We show that this model is not plausible for \is{most} known SNRs and \is{can only marginally explain the gamma-ray emission from the least luminous hadronic-dominated SNR IC~443, but even in this case the model is in tension with the observed radio emission.} In general \is{this scenario} could be feasible only in very specific cases. Moreover, the feasibility of the model could be directly checked through radio observations as GCR electrons should also be compressed and leave a characteristic \is{curvature} in the synchrotron spectrum at radio wavelengths \is{corresponding to the break in the spectrum of GCR electrons. Such a curvature is not observed for evolved SNRs which exhibit featureless power-law spectra across a wide range of frequencies. To decisively clarify this observations at $\sim{100}$~GHz would be necessary}. We find, however, that synchrotron radiation from compressed electrons would \is{be very low contributing} at most $\sim10\,\%$ to the observed radio-luminosity. \rb{The need for a large number of freshly accelerated electrons makes it unlikely that protons get not freshly accelerated and likewise exceeding the contribution of the compressed CRs.} Finally, we show that the confinement of high-energy protons could be a problem in the compressed shell which farther strongly disfavours this scenario.

\section*{Acknowledgements}
Robert Brose acknowledges funding from the Irish Research Council under the Government of Ireland Postdoctoral Fellowship program. Iurii Sushch acknowledges support by the National Research Foundation of South Africa (Grant Number 132276).



\section*{Data availability}
The data underlying this article can be shared on reasonable request to the corresponding author.

\bibliographystyle{mnras}
\bibliography{ref}

\appendix
\section{Models for the IC~443 SNR}
\label{app:models}

We construct two models specifically for the IC~443 SNR based on its morphology and size. \is{For both models we utilize} 
the two-shell idea adopted from \citet{2006ApJ...649..258T} and references therein. We assume that the SNR shell can be approximated by two hemispheres of radii $R_\mathrm{A} = 7$~pc and $R_\mathrm{B} = 11$~pc. \is{For the first model we assume uniform dense media in both hemispheres while for the second model we consider a complex clumpy environment.}

\is{{\bf UNIFORM DENSE MEDIA.} In this model} the densities of media \is{in both hemispheres} are \is{assumed to be} constant. \is{The age of IC~443 is very uncertain ranging from $\sim1000$ yr implied from X-ray observations \citep{1988ApJ...335..215P,1992PASJ...44..303W} to $\sim30$~kyr suggested by the proper motion of the compact object potentially associated with the SNR.}
We estimate the highest possible values for densities $n_\mathrm{A} = 50\,\mathrm{cm}^{-3}$ and $n_\mathrm{B} = 9\,\mathrm{cm}^{-3}$ following analytic approximations from \citet{1988ApJ...334..252C} and adopting the upper limit on the discussed range of possible ages for the remnant of $t_\mathrm{age} = 30$~kyr. \is{These estimates are consistent both with analytic treatment \citep{1999ApJ...511..798C} and numeric simulations of the expansion of IC~443 into a molecular cloud \citep{2019MNRAS.482.1602Z}.}
We also assume that both shells are radiative and the swept-up material as well as CRs are compressed with the compression ratio of $\chi = 30$. The expected gamma-ray luminosity from compressed GCRs fails to explain the observed gamma-ray emission (\is{top panel in Fig.} \ref{fig:gammaray_lum}). \is{Note, in this model we calculate the gamma-ray luminosity from the whole SNR, although the gamma-ray emission is detected only from the hemisphere A.}

\rb{To calculate the radio emission, we assume a magnetic field strength we follow Eq.~\ref{eq:B_Crutcher} for the upstream field of both hemispheres. The field is then compressed in the downstream, resulting in a higher field-strength. Again, the chosen value for $B$ has to be considered as an upper limit and so is the derived radio-flux displayed in the top panel of Fig. \ref{fig:W44_Radio}.}

\is{{\bf COMPLEX CLUMPY ENVIRONMENT.}} 
\is{IC 443 interacts with a very complex environment that consists of molecular and atomic clouds \citep[see e.g.][]{2001ApJ...547..885R, 2014ApJ...788..122S}. \citet{2021A&A...649A..14U} presented a detailed 3D hydrodynamic model for IC~443 which accounts for interaction with the atomic cloud in the northern hemisphere A and with the toroidal molecular cloud that encircles the remnant between two hemispheres and is responsible for the bright southern ridge. The best fit of their model to the multiwavelength observational data is obtained for the density of the atomic cloud to be about $300\,\mathrm{cm}^{-3}$ and the density of the molecular cloud to be about $300\,\mathrm{cm}^{-3}$ which is in good agreement with numbers obrained by \citet{2001ApJ...547..885R} from near-infrared observations, $10-1000$ and $\sim10^4\,\mathrm{cm}^{-3}$ respectively. Initially the remnant is assumed to propagate into the intercloud medium with the density of $0.2\,\mathrm{cm}^{-3}$ and then at some point starts to interact with clouds in the northern hemispheres A, while the southern hemisphere B continues to evolve in the intercloud medium. The age of the remnant is found to be around 8 kyr. Although it is clear that a large fraction of the shock surface interacts with very dense clouds at different times, it is also evident from 3D hydrodynamic simulations that the shock does not propagate strongly into the cloud, but rather gets contained by the cloud \citep{2021A&A...649A..14U}. Hence, the amount of the cloud material in the compressed shell behind the shock should be rather small and the expected gamma-ray emission from compressed GCRs should be negligible.}

\is{To construct our second model we roughly follow the setup of \citep{2021A&A...649A..14U} but modify it introducing clumpiness to make the scenario of CR compression more optimistic. We assume that the northern hemisphere A throughout its evolution was propagating into the clumpy medium with the density of clumps $n_\mathrm{c} = 1000\,\mathrm{cm}^{-3}$ and density of the intercloud medium of $n_\mathrm{ic} = 10\,\mathrm{cm}^{-3}$. This description nicely follows measurements by \citet{2001ApJ...547..885R}. The expansion of the shock in the clumpy environment can be described by the analytic solution proposed by \citet{1991ApJ...373..543W} which is in good agreement with numeric simulations \citep{2017ApJ...846...77S}. This solution implies that the shock radius can be estimated by simple scaling of the Sedov-Taylor radius
\begin{equation}
    R = R_\mathrm{ST} \left(\frac{K}{1.528}\right)^{1/5}
\end{equation}
with the parameter $K$ which depends on the cloud-to-intercloud mass ratio $C$ and evaporation time scale $\tau = t_\mathrm{evap}/t_\mathrm{age}$. Numeric simulations by \citet{2017ApJ...846...77S} for the cloud density of $25\,\mathrm{cm}^{-3}$ imply $\tau\approx5$ roughly constant throughout evolution. Given that the $t_\mathrm{evap}\propto n_\mathrm{c}$ \citep{2017ApJ...846...77S}, $n_\mathrm{c} = 1000\,\mathrm{cm}^{-3}$ yields $\tau \sim 200$. We also set $C=10$ that corresponds to the filling factor of $\phi = 0.091$ for the assumed densities
\begin{equation}
    \phi = \frac{C}{\frac{n_\mathrm{c}}{n_\mathrm{ic}}+C}. 
\end{equation}}
\is{For $C=10$ and $\tau = 200$ the \citep{1991ApJ...373..543W} solution basically converges to the Sedov-Taylor solution and the shock should reach $R_\mathrm{A} = 7$~pc in $t_\mathrm{age}=7400$ yr. This age implies that the density of the medium in the hemisphere B is $1.4\,\mathrm{cm}^{-3}$.

For the molecular cloud we assume the density of $n_\mathrm{m}=10^4\,\mathrm{cm}^{-3}$ and the velocity of the shock interacting with the cloud of $v_\mathrm{m} = 30$~km/s in agreement with \citet{2001ApJ...547..885R}. More recent work by \citet{2022MNRAS.511..953C} reports similar measurements based on ESO–ARO Public Spectroscopic Survey SHREC with the shocked gas density of $\geq10^5\,\mathrm{cm}^{-3}$, a factor of $>10$ higher than the pre-shock density, and the shock velocity of $\sim23$~km/s. Further, to remain on the optimistic side, we assume that the distance the the shock propagates into the cloud is 
\begin{equation}
    \Delta R_\mathrm{m} = v_\mathrm{m} t_\mathrm{age} = 0.2\,\mathrm{pc},
\end{equation}
and that the cloud covers $\omega = 0.1$ of the surface of the hemisphere A. 
Corresponding volumes of the material in the clumpy medium and in the molecular cloud that the shock interacted with can be written as 
\begin{align}
    V_\mathrm{c} &= (1-\omega)\frac{2}{3}\pi R_\mathrm{A}^3 \\
    V_\mathrm{m} &=  \omega 2 \pi R_\mathrm{A}^2 \Delta R_\mathrm{m}
\end{align}
and the total volume of the crushed shell thus 
\begin{equation}
    V_\mathrm{shell} = \chi (V_\mathrm{c} + V_\mathrm{m})
\end{equation}
where the compression ratio is assumed to be the same and $\chi = 30$. The hemisphere B is completely ignored \rb{for the gamma-ray emission as} the contribution is negligible and because no significant gamma-ray emission is detected from that region. The computed total gamma-ray luminosity as well as luminosities of separate components are shown in  the bottom panel of Fig. \ref{fig:gammaray_lum}. It can be seen that this model can roughly explain the observed luminosity from IC 443, but it should be stressed again that the model was purposefully constructed to be too optimistic.}

\rb{For the radio-emission, we consider both hemispheres as the larger volume of hemisphere B yields to a significant amount of emission in face of the  comparable field-strength in both hemispheres. In addition to the contribution from the inter-cloud medium of both hemispheres, with a field-strength of $B=10\,\mu$G, we assume that the field reaches the limits given in \cite{2010ApJ...725..466C} for the clumps of hemisphere B and the toroidal cloud (see Eq.~\ref{eq:B_Crutcher}).

The radio emission from the hemispheres A and B as well as from the cloud is shown in the bottom panel of Fig. \ref{fig:W44_Radio}. As a consequence of the compressed field in the clumps of hemisphere A, the radio emission from the clumps is comparable to the emission of the inter-clump medium from that hemisphere. However, the strongest emission has to be expected from hemisphere B in that model on account of the higher total number of electrons in that region.}



\bsp	
\label{lastpage}
\end{document}